\documentclass[twocolumn,superscriptaddress,showpacs,preprintnumbers,amsmath,amssymb,prl]{revtex4}
\usepackage{graphicx}
\usepackage{dcolumn}
\usepackage{bm}
\usepackage{longtable}%
\usepackage{dcolumn}
\usepackage{colordvi}\usepackage[usenames]{color}
\def\etal{{\it et al.}}
\newcommand{\header}[1]{{\em#1.---}}
\def\Titech{Department of Physics, Tokyo Institute of Technology, 2-12-1 O-Okayama, Meguro, Tokyo 152-8551, Japan}
\def\LPC{LPC Caen, Normandie Universit\'e, ENSICAEN, Universit\'e de Caen, CNRS/IN2P3, F-14050, Caen, France}
\def\SNU{Department of Physics and Astronomy, Seoul National University, 599 Gwanak, Seoul 151-742, Republic of Korea}
\def\RIKEN{RIKEN Nishina Center, Hirosawa 2-1, Wako, Saitama 351-0198, Japan}
\def\TMD{Institut f\"ur Kernphysik, Technische Universit\"{a}t Darmstadt, D-64289 Darmstadt, Germany}
\def\TOHOKU{Department of Physics, Tohoku University, Miyagi 980-8578, Japan}
\def\RIKKYO{Departiment of Physics, Rikkyo University, Toshima, Tokyo 171-8501, Japan} 
\def\KYOTO{Department of Physics, Kyoto University, Kyoto 606-8502, Japan}
\def\GANIL{GANIL, CEA/DRF-CNRS/IN2P3, F-14076 Caen Cedex 5, France}
\def\EMMIGSI{ExtreMe Matter Institute EMMI and Research Division, GSI Helmholtzzentrum f\"ur Schwerionenforschung GmbH, D-64291 Darmstadt, Germany}
\def\YORK{Department of Physics, University of York, Heslington, York YO10 5DD, United Kingdom}
\def\ORSAY{Institut de Physique Nucl\'eaire, IN2P3-CNRS, Universit\'e Paris-Sud, Universit\'e Paris-Saclay, F-91406 Orsay Cedex, France}
\def\MSU{NSCL/FRIB Laboratory, Michigan State University, East Lansing, Michigan 48824, USA}
\def\CHINA{School of Physics, Peking University, Beijing 100871, China}
\begin{document}
\title{{\boldmath{}First observation of $^{20}$B and $^{21}$B}}
\author{S.~Leblond}\affiliation{\LPC}
\author{F.M.~Marqu\'es}\affiliation{\LPC}
\author{J.~Gibelin}\affiliation{\LPC}
\author{N.A.~Orr}\affiliation{\LPC}
\author{Y.~Kondo}\affiliation{\Titech}
\author{T.~Nakamura}\affiliation{\Titech}
\author{J.~Bonnard}\affiliation{\ORSAY}
\author{N.~Michel}\affiliation{\MSU}\affiliation{\CHINA}
\author{N.L.~Achouri}\affiliation{\LPC}
\author{T.~Aumann}\affiliation{\TMD}\affiliation{\EMMIGSI}
\author{H.~Baba}\affiliation{\RIKEN}
\author{F.~Delaunay}\affiliation{\LPC}
\author{Q.~Deshayes}\affiliation{\LPC}
\author{P.~Doornenbal}\affiliation{\RIKEN}
\author{N.~Fukuda}\affiliation{\RIKEN}
\author{J.W.~Hwang}\affiliation{\SNU}
\author{N.~Inabe}\affiliation{\RIKEN}
\author{T.~Isobe}\affiliation{\RIKEN}
\author{D.~Kameda}\affiliation{\RIKEN}
\author{D.~Kanno}\affiliation{\Titech}
\author{S.~Kim}\affiliation{\SNU}
\author{N.~Kobayashi}\affiliation{\Titech}
\author{T.~Kobayashi}\affiliation{\TOHOKU}
\author{T.~Kubo}\affiliation{\RIKEN}
\author{J.~Lee}\affiliation{\RIKEN}
\author{R.~Minakata}\affiliation{\Titech}
\author{T.~Motobayashi}\affiliation{\RIKEN}
\author{D.~Murai}\affiliation{\RIKKYO}
\author{T.~Murakami}\affiliation{\KYOTO}
\author{K.~Muto}\affiliation{\TOHOKU}
\author{T.~Nakashima}\affiliation{\Titech}
\author{N.~Nakatsuka}\affiliation{\KYOTO}
\author{A.~Navin}\affiliation{\GANIL}
\author{S.~Nishi}\affiliation{\Titech}
\author{S.~Ogoshi}\affiliation{\Titech}
\author{H.~Otsu}\affiliation{\RIKEN}
\author{H.~Sato}\affiliation{\RIKEN}
\author{Y.~Satou}\affiliation{\SNU}
\author{Y.~Shimizu}\affiliation{\RIKEN}
\author{H.~Suzuki}\affiliation{\RIKEN}
\author{K.~Takahashi}\affiliation{\TOHOKU}
\author{H.~Takeda}\affiliation{\RIKEN}
\author{S.~Takeuchi}\affiliation{\RIKEN}
\author{R.~Tanaka}\affiliation{\Titech}
\author{Y.~Togano}\affiliation{\Titech}\affiliation{\EMMIGSI}
\author{A.G.~Tuff}\affiliation{\YORK}
\author{M.~Vandebrouck}\affiliation{\ORSAY}
\author{K.~Yoneda}\affiliation{\RIKEN}
\date{\today}%
\begin{abstract}
 The most neutron-rich boron isotopes $^{20}$B and $^{21}$B have been observed for the first time following proton
 removal from $^{22}$N and $^{22}$C at energies around 230~MeV/nucleon. Both nuclei were found to exist as resonances
 which were detected through their decay into $^{19}$B and one or two neutrons. Two-proton removal from $^{22}$N
 populated a prominent resonance-like structure in $^{20}$B at around 2.5~MeV above the one-neutron decay threshold,
 which is interpreted as arising from the closely spaced $1^-,2^-$ ground-state doublet predicted by the shell model.
 In the case of proton removal from $^{22}$C, the $^{19}$B plus one- and two-neutron channels were consistent with the
 population of a resonance in $^{21}$B $2.47\pm0.19$~MeV above the two-neutron decay threshold, which is found to
 exhibit direct two-neutron decay. The ground-state mass excesses determined for $^{20,21}$B are found to be in
 agreement with mass surface extrapolations derived within the latest atomic-mass evaluations.
\end{abstract}
\maketitle

\header{Introduction}
 The advent of dedicated radioactive-beam facilities has provided for a rather complete mapping of the nuclear
 landscape up to mass number $\sim30$ \cite{Nakamura}. As such it is now well established that the text-book shell
 structure of the nucleus, that translates into an enhanced stability for systems with ``magic'' numbers of protons
 ($Z$) and/or neutrons ($N$) of 2, 8, 20...\ is modified as the limits of particle stability, or driplines, are
 approached (see, for example, Ref.~\cite{Dripline}). Significantly, these changes in shell structure, which have
 been attributed to a number of different mechanisms, including most recently and intriguingly the effects of
 three-body forces \cite{Otsuka10}, influence the location of the dripline itself.

\begin{figure}
 \begin{center}
 \includegraphics[width=.48\textwidth]{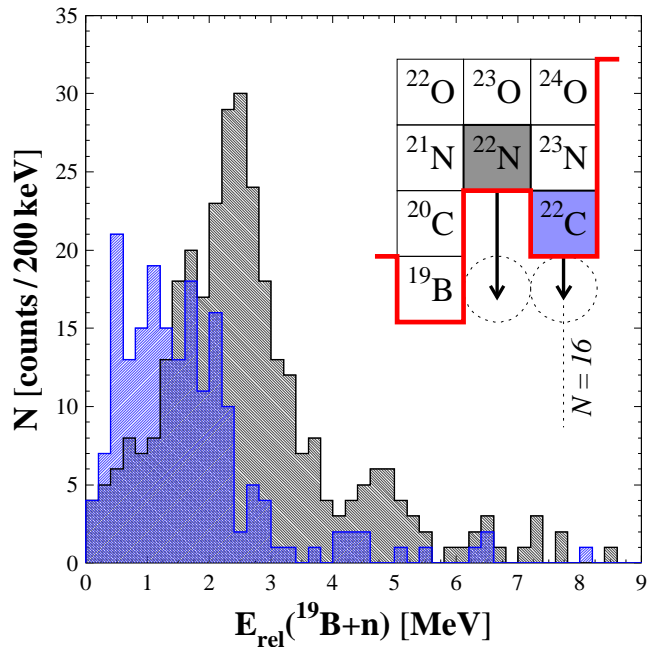}
 \caption{Relative energy spectrum of $^{19}$B+$n$ events following proton-removal from $^{22}$N (gray)
 and $^{22}$C (blue histogram). The red line in the inset delineates the neutron dripline.} \label{f:Eds}
 \end{center}
\end{figure}

 In the naive shell-model picture, neutron numbers between 8 and 20 correspond to the filling of the $sd$-shell
 neutron single-particle orbitals ($\nu 0d_{5/2}$, $\nu 1s_{1/2}$, $\nu 0d_{3/2}$). Approaching the neutron dripline,
 the energies of these orbitals evolve, leading for example to the disappearance of the $N=20$ magic number for
 $Z=10$--12 (the so-called ``Island of Inversion'' \cite{IoI}) and to the appearance of new shell closures at $N=14$
 and 16 in the oxygen isotopes \cite{Stan04,Hoffman,BAB-WAR}. In this respect, the most neutron-rich boron isotopes,
 which lie below doubly-magic $^{22,24}$O and straddle the neutron dripline, are of considerable interest
 (Fig.~\ref{f:Eds}, inset) and, significantly, are now coming within the range of sophisticated {\em ab initio}
 models \cite{abnit} and approaches that treat explicitly the continuum \cite{GSM}.
 More generally, the boron isotopic chain exhibits a number of exotic structures: from the proton halo of $^8$B
 \cite{Boron8}, through the unbound threshold states of $^{16,18}$B \cite{JL,Spyrou10}, to the two-neutron halo of
 $^{17}$B and the two/four neutron halo/skin of $^{19}$B \cite{Suzuki99}.  
 
 This Letter reports on the first observation of the neutron-unbound nuclei $^{20}$B and $^{21}$B \cite{Ozawa2003},
 populated through high-energy proton removal and reconstructed using invariant-mass spectroscopy. These measurements,
 at the limits of present capabilities, provide for the first experimental mass determinations for both isotopes.
 In addition, evidence is presented showing that $^{21}$B decays by direct two-neutron emission. Finally, a comparison
 with the predictions of shell-model calculations is discussed and provisional spin-parity assignments provided for
 the levels observed.  
  
\header{Experiment}
 The experiment was performed at the Radioactive Isotope Beam Factory (RIBF) of the RIKEN Nishina Center,
 as part of an experimental campaign investigating the structure of light neutron-rich nuclei beyond the dripline
 (see, for example, Refs.~\cite{Kondo16,Togano16}). Secondary beams of $^{22}$N and $^{22}$C were produced by
 fragmentation of a 345~MeV/nucleon $^{48}$Ca primary beam incident on a 20~mm thick beryllium target, and were
 separated using the BigRIPS fragment separator \cite{Kubo03}. The different isotopes present in the secondary beams
 were identified via the measurement of their energy loss, time of flight and magnetic rigidity, and transported to
 the object point of the SAMURAI spectrometer \cite{Koba13}, where a 1.8~g/cm$^2$ carbon reaction target was located.
 The beam particles were tracked onto the target using two drift chambers. The energies at target mid-point and average
 intensities of the $^{22}$N and $^{22}$C beams were, respectively, 225 and 233~MeV/nucleon, and 6600 and 6~pps.

 The beam-velocity reaction products were detected in the forward direction using the SAMURAI setup including the
 NEBULA neutron array \cite{Kondo15}, placed some 11~m downstream of the target. The SAMURAI superconducting dipole
 magnet \cite{Sato-SAMURAI}, with a central field of 3~T, provided for the momentum analysis of the charged fragments.
 The dipole gap was kept under vacuum using a chamber equipped with thin exit windows \cite{Shimizu}
 so as to reduce to a minimum the amount of material encountered by both the fragments and neutrons.
 Drift chambers at the entrance and exit of the magnet allowed the determination of their trajectories and
 magnetic rigidity \cite{Koba13}. This information, combined with the energy loss and time of flight measured
 using a 16-element plastic hodoscope, provided for the identification of the projectile-like fragments.
 The neutron momenta were derived from the time of flight, with respect to a thin plastic start detector
 positioned just upstream of the target, and the hit position measured with the 120 plastic scintillator modules
 (12$\times$12$\times$180~cm$^3$) of the NEBULA array.
 
\header{Results}
 The relative energy ($E_{\rm{rel}}$) of the unbound boron isotopes was reconstructed from the momenta of the $^{19}$B
 fragment and neutron(s) as the invariant mass of the $^{19}$B+$xn$ system minus the masses of the constituents.
 It should be noted that $^{19}$B, owing to its extremely weakly-bound character
 (two-neutron separation energy of $0.14\pm0.39$~MeV \cite{Gaud12}), has no bound excited
 states and thus $E_{\rm{rel}}$ reflects directly the energy above the decay threshold.
 The spectra reconstructed using $^{19}$B+$n$ events from reactions induced by the $^{22}$N and $^{22}$C beams
 are shown in Fig.~\ref{f:Eds}, and exhibit significant differences. In particular, while two-proton removal from
 $^{22}$N populates a clear resonance-like structure around 2--3~MeV, proton removal from $^{22}$C leads to a very
 broad distribution confined to energies below $\sim2.5$~MeV.
 
 None of these features can be attributed to the response function of the setup, as it varies smoothly with
 $E_{\rm{rel}}$ (see, for example, Fig.~1 of Ref.~\cite{Kondo16}). In order to deduce the character of any resonances
 in $^{20,21}$B, the spectra were described using single-level R-matrix line-shapes \cite{BW} which were used as the
 input for a complete simulation of the setup (including the secondary-beam characteristics, the reaction, and the
 detector resolutions and acceptances) together with a non-resonant component. The resolution ({\sc{fwhm}}) in the
 reconstructed $E_{\rm{rel}}$ was dominated by the NEBULA hit position determination and timing resolution, and
 varied as $\sim0.4\sqrt{E_{\rm{rel}}}$~MeV.

\begin{figure}
 \begin{center}
 \includegraphics[width=.48\textwidth]{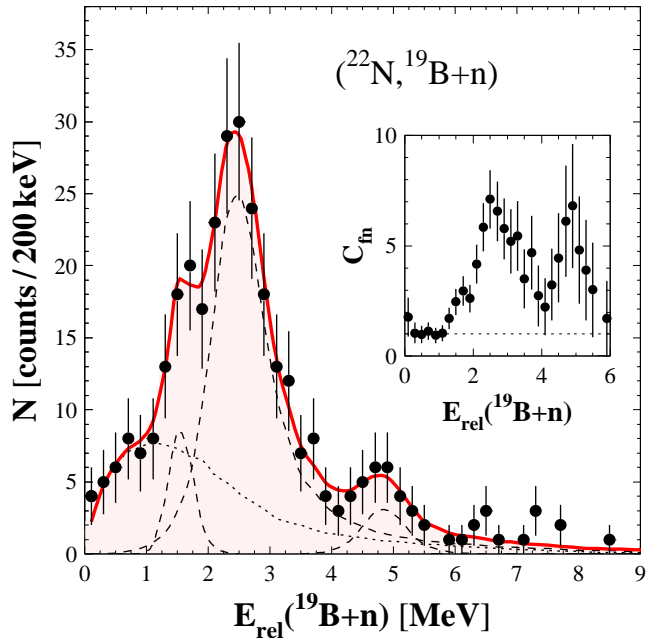}
 \caption{Relative-energy spectrum of $^{19}$B+$n$ events following two-proton removal from $^{22}$N. The red line
 corresponds to the best fit ($\chi^2/{\rm{ndf}}=0.33$), including the non-resonant continuum (dotted) and $^{20}$B
 resonances at 1.56, 2.50 and 4.86~MeV (dashed lines). The inset shows the fragment-$n$ correlation function $C_{fn}$
 (see text).} \label{f:E20}
 \end{center}
\end{figure}

\begin{figure}
 \begin{center}
 \includegraphics[width=.48\textwidth]{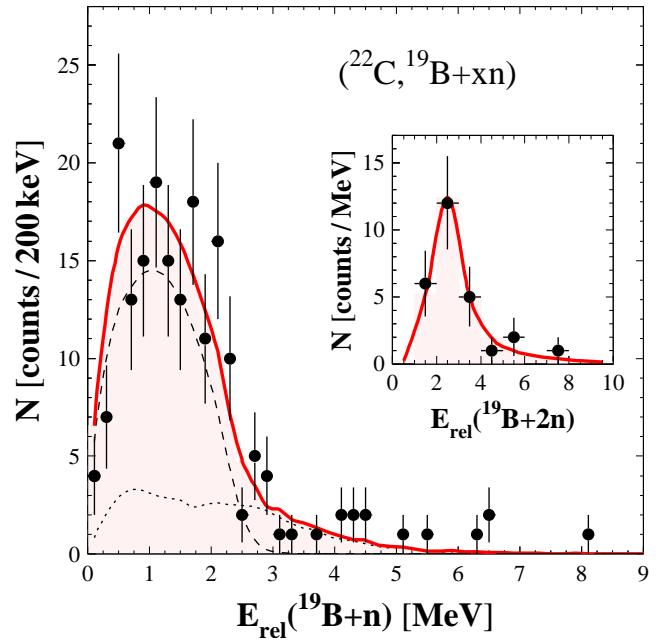}
 \caption{Relative-energy spectrum of $^{19}$B+$n$ following one-proton removal from $^{22}$C. The red line
 corresponds to the best fit ($\chi^2/{\rm{ndf}}=1.2$), including the non-resonant continuum (dotted) and the
 phase-space decay from a $^{21}$B resonance at 2.47~MeV (dashed line). The inset shows the spectrum of
 $^{19}$B+$2n$ events, with the best fit (red line) for a $^{21}$B resonance at 2.4~MeV.} \label{f:E21}
 \end{center}
\end{figure}

 The shape of the non-resonant continuum was deduced for each reaction channel by mixing the measured $^{19}$B-$n$
 pairs following the procedure described in Ref.~\cite{MIX}. Importantly, the uncorrelated distribution so obtained
 does not require any {\em a priori} parameterizations and incorporates explicitly the effects of the experimental
 response function. As such, it may be compared directly with the measured distribution in order to identify features
 arising from the decay of unbound states \cite{Giacomo}. As may be seen in Figs.~\ref{f:E20} and \ref{f:E21}, the
 non-resonant distributions for the $^{19}$B+$n$ events from the $^{22}$N and $^{22}$C beams clearly cannot 
 account for the prominent structures in either case.
 
 Turning first to the results for two-proton removal from $^{22}$N, the inset of Fig.~\ref{f:E20} displays the
 correlation function obtained as the ratio of the data and the uncorrelated non-resonant distribution \cite{MIX}. 
 Importantly, in addition to displaying more clearly the presence of a peak at about 5~MeV, the region below
 1~MeV shows no resonant signal.
 In terms of resonances in $^{20}$B, only decays to the $^{19}$B ground state by $\ell=2$ neutron emission are
 expected to be observable \footnote{Decay by $\ell=0$ neutron emission for all but threshold states will result in
 extremely broad structures.}. In particular, the single-particle width for a $d$-wave resonance at 2.5~MeV is,
 assuming a standard Woods-Saxon potential, $\sim1.3$~MeV. A fit in terms of a single prominent resonance at about
 2.5~MeV and a weaker high-lying one (plus the non-resonant continuum) provides for a good description of the spectrum,
 with the energy and width of the former $E_r=2.44\pm0.09$~MeV and $\Gamma=1.2\pm0.4$~MeV \cite{Supp-url}.
 Such a width suggests that the spectroscopic factor for the decay to the $^{19}$B ground state is large.
 Simple considerations, however, suggest that the lowest-lying levels of $^{20}$B will be a $1^-,2^-$ doublet
 arising from the coupling of a $0p_{3/2}$ proton with a $1s_{1/2}$ neutron, and that the strong peak observed here
 may well result from the population of both states. This point, and the related fit shown in Fig.~\ref{f:E20},
 is addressed in the discussion below in the light of shell-model calculations.

 In the case of single-proton removal from $^{22}$C (Fig.~\ref{f:E21}), the $^{19}$B+$n$ channel does not exhibit
 any clear peaks arising from resonances in $^{20}$B, but rather a ``plateau-like'' distribution, modulated by the
 experimental response function, reminiscent of the direct phase-space decay of a three-body resonance
 \cite{Supp-url}, in this case $^{21}$B. Despite the reduced two-neutron detection efficiency,
 the relative energy spectrum of $^{19}$B+$2n$ events, after applying cross-talk rejection conditions \cite{crosstalk},
 could be reconstructed as shown in the inset of Fig.~\ref{f:E21}. It displays clearly resonance-like strength
 in the region around 2.5~MeV. Using a simple Breit-Wigner line-shape with an energy-dependent width, the best fit
 was for a $^{21}$B resonance at $E_r=2.4\pm0.4$~MeV with $\Gamma<3$~MeV. While the very limited statistics precluded
 the construction of the event-mixed three-body non-resonant continuum, the influence of such a distribution is
 expected to be less than the quoted uncertainties.

 Having established that the reactions induced by the $^{22}$C beam populate a resonance-like structure in $^{21}$B,
 a more precise energy and width may be derived from the higher statistics two-body ($^{19}$B+$n$) data set,
 as was the case in the study of $^{26}$O \cite{Kondo16,GSI-26O}. 
 The $^{19}$B+$n$ spectrum was fitted with a combination of the uncorrelated distribution derived from event mixing
 and simulated events arising from the decay of a resonance in $^{21}$B.
 The latter was assumed to occur by three-body phase space into $^{19}$B+$n$+$n$, and $E_{\rm{rel}}$ was reconstructed
 between the fragment and the neutron with the shortest time of flight (the procedure employed in the treatment of the
 data). The energy and width of the $^{21}$B resonance are sensitive to the location and slope, respectively, of the
 higher-energy edge of the $^{19}$B+$n$ distribution \cite{Supp-url}.

 The best fit, shown in Fig.~\ref{f:E21}, is for a resonance in $^{21}$B with $E_r=2.47\pm0.19$~MeV and
 $\Gamma<0.6$~MeV. The errors include a systematic uncertainty derived from other direct-decay modes, in which
 the $n$-$n$ interaction modifies the three-body phase space \cite{Supp-url} following the formalism of
 Ref.~\cite{Dalitz}. Given the very good description of the $^{19}$B+$n$ events from the $^{22}$C beam, any
 contribution from sequential decay through (and/or direct population of) $^{20}$B must be small ($\sim10$\%)
 \cite{Supp-url}. This is consistent with the resonances found here in $^{20}$B being at similar energy or higher
 than in $^{21}$B, providing little or no opportunity for sequential decay to occur.
 As such, $^{21}$B may be considered a new case of direct two-neutron decay.

\header{Discussion}
 In the following the present results are discussed in the light of shell-model calculations (SM), that
 were undertaken \cite{SM} in the full $psd$ model space (that is, comprising the $0p_{3/2}$, $0p_{1/2}$, $0d_{5/2}$,
 $1s_{1/2}$ and $0d_{3/2}$ single-particle orbits) for protons and neutrons using the monopole-based universal
 interaction YSOX \cite{Yuan12}, which successfully reproduces the location of the neutron dripline for carbon and
 oxygen. The spurious center-of-mass contributions were removed using the Lawson prescription. Configurations
 corresponding to up to five particle-hole excitations ($5\hbar\omega$) were included in the many-body space, but only
 small differences in the energies were observed with the $3\hbar\omega$ approximation used to design the interaction.

\begin{figure}
 \begin{center}
 \includegraphics[width=.48\textwidth]{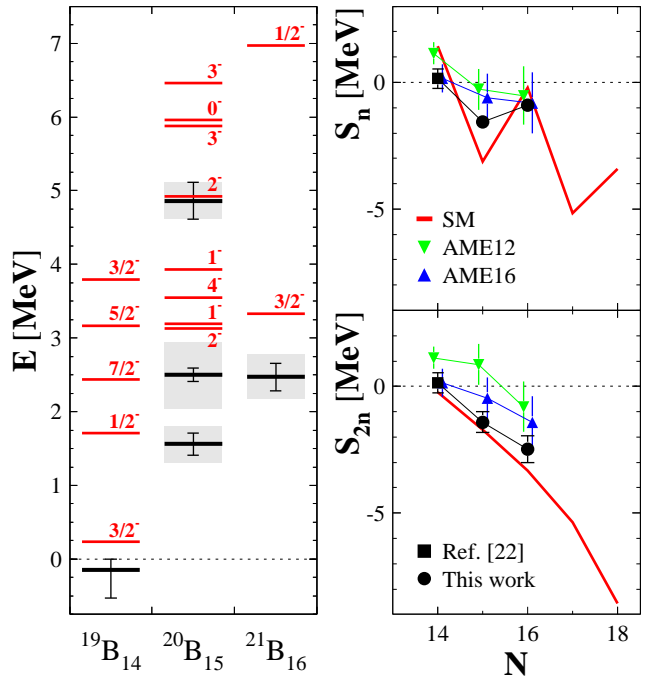}
 \caption{Left: experimental (black) and shell-model (red) levels with respect to the $2n$ ($^{19}$B),
 $1n$ ($^{20}$B) and $2n$ ($^{21}$B) thresholds. The experimental uncertainties on the energies are shown 
(taking into account the particle stability of $^{19}$B), as well as the widths of the resonances (gray boxes).
 Right: one- and two-neutron separation energies for experimental data (black), shell-model predictions (red),
 and mass-surface extrapolations (green and blue points).} \label{f:AME}
 \end{center}
\end{figure}

 As alluded to above, the low-lying spectrum of $^{20}$B ($N=15$) should exhibit a series of states arising from
 the coupling of the odd valence neutron with a proton hole in the $0p_{3/2}$ orbit. While the calculation of
 two-proton removal reaction cross-sections is complex and beyond the scope of this work, the observation that
 single-proton removal from $^{22}$N (which exhibits a strong $1s_{1/2}$ valence neutron configuration) populates
 almost exclusively the $1/2^+$ ground state of $^{21}$C \cite{Sylvain} suggests that removal of a second proton
 should favor population of a $1^-,2^-$ doublet, one member of which would be expected to be the $^{20}$B ground
 state. Indeed, as seen in Fig.~\ref{f:AME}, the SM predicts these to be the lowest-lying levels with a very
 small separation. In addition, both states are predicted to exhibit $d$-wave neutron decay branches
 (Table~\ref{t:C2S}), and the corresponding decay widths will thus be much less than the single-particle value
 of $\sim$1.3~MeV noted earlier.

 In the light of these considerations, the $^{19}$B+$n$ relative energy spectrum was fitted assuming the structure
 at around 2.5~MeV to be composed of two closely spaced narrower $d$-wave resonances. As shown in Fig.~\ref{f:E20},
 the inclusion of such a doublet, in addition to the high-lying resonance and the non-resonant continuum, allows the
 spectrum to be very well reproduced
 \footnote{Statistically, the improvement in the goodness of the fit compared to assuming a single resonance
 \cite{Supp-url} is not significant, from $\chi^2/{\rm{ndf}}=0.45$ to 0.33.}.
 The best fit parameters for the three resonances were: $E_r=1.56\pm0.15$, $2.50\pm0.09$, and $4.86\pm0.25$~MeV;
 and $\Gamma<0.5$, $0.9\pm0.3$, and $<0.5$~MeV. A comparison with the shell-model calculations is shown in
 Fig.~\ref{f:AME}, whereby the energy with respect to the first particle-emission threshold is plotted.
 Given that the total binding energies are $\sim60$~MeV, the energies of the predicted ground states
 and lowest-lying levels observed here are in reasonable accord. 
 
 In terms of excited states of $^{20}$B, taking into account the underbinding of the SM calculations
 ($\sim1.5$~MeV), one may speculate that the weaker peak observed at 4.86~MeV ($E_x=3.30\pm0.29$~MeV) could
 correspond to the $0_1^-$ and/or $3_2^-$ levels (Fig.~\ref{f:AME}), which are predicted to exhibit significant
 spectroscopic strength for neutron decay to $^{19}$B ground state (Table~\ref{t:C2S}). It is interesting to note
 that the very weakly-bound character of $^{19}$B means that all the $^{20}$B levels observed here are energetically
 permitted to decay via $3n$ emission to $^{17}$B.

 Turning to $^{21}$B ($N=16$), the SM predicts a $3/2^-$ ground state formed by the $0p_{3/2}$ proton hole and
 expected to be the only state populated with any observable strength following proton removal from $^{22}$C.
 Preliminary estimates made using the Gamow Shell Model \cite{GSM} suggest a $^{21}$B ground state (unbound by
 about 1.7~MeV) with a width of $\Gamma\sim130$~keV \cite{Nicolas}, consistent with our upper limit of 600~keV.
 
 In terms of the $N=16$ shell closure, the SM predicts a rather high-lying first excited state ($1/2^-$) in $^{21}$B,
 although with an excitation energy $E_x=3.6$~MeV lower than found experimentally in $^{24}$O (4.7~MeV
 \cite{Hoffman,Tshoo12}) and $^{23}$N (4.1~MeV~\footnote{The observation of Ref.~\cite{Jones17} combined with the
 most recent evaluation of the single-neutron separation energy, $S_n=3.12\pm0.47$~MeV \cite{AME16}.}),
 and than predicted by these calculations in $^{22}$C (5.0~MeV).
 Given that the SM predicts the first excited state of $^{21}$B to have a strength some ten times less than the
 ground state in proton removal from $^{22}$C, its non-observation here is not surprising. 

\begin{table}
 \caption{SM predictions for the spectroscopic factors for the decay
  of $^{20}$B levels (Fig.~\ref{f:AME}) to the $^{19}$B ground state.} \label{t:C2S}
 \begin{tabular}
  {p{\dimexpr.25\linewidth-2\tabcolsep}p{\dimexpr.25\linewidth-2\tabcolsep}
   p{\dimexpr.25\linewidth-2\tabcolsep}p{\dimexpr.25\linewidth-2\tabcolsep}} \hline\hline
 $E$ (MeV)	& $J^\pi$	& $\ell_n$	& $C^2S$	\\ \hline
 3.13		& $2_1^-$	& 0			& 0.21		\\
			&			& 2			& 0.16		\\
 3.19		& $1_1^-$	& 0			& 0.09		\\
			&			& 2			& 0.52		\\
 3.55		& $4_1^-$	& 2			& 0.30		\\
 3.93		& $1_2^-$	& 0			& 0.07		\\
			&			& 2			& 0.17		\\
 4.93		& $2_2^-$	& 2			& 0.05		\\
 5.88		& $3_1^-$	& 2			& 0.10		\\
 5.96		& $0_1^-$	& 2			& 0.43		\\
 6.46		& $3_2^-$	& 2			& 0.70		\\ \hline\hline           
 \end{tabular}
\end{table}

\begin{table}
 \caption{Experimental mass excesses (MeV) of the heaviest boron isotopes (present work and Ref.~\cite{Gaud12})
 compared to the most recent atomic-mass evaluations \cite{AME16,AME12}.} \label{t:Mex}
 \begin{tabular}{p{\dimexpr.19\linewidth-2\tabcolsep}p{\dimexpr.27\linewidth-2\tabcolsep}
                 p{\dimexpr.27\linewidth-2\tabcolsep}p{\dimexpr.27\linewidth-2\tabcolsep}} \hline\hline
 Isotope	& AME12	\cite{AME12}& AME16 \cite{AME16}& Experiment		\\ \hline
 $^{19}$B	& 58.78$\pm$0.40	& 59.77$\pm$0.53	& 59.77$\pm$0.35	\\ 
 $^{20}$B	& 67.13$\pm$0.70	& 68.45$\pm$0.80	& 69.40$\pm$0.38	\\
 $^{21}$B	& 75.72$\pm$0.90	& 77.33$\pm$0.90	& 78.38$\pm$0.40	\\ \hline\hline
 \end{tabular}
\end{table}

 Assuming that the lowest-lying levels observed here correspond to the ground states of $^{20,21}$B, the resonance
 energies, in combination with the $^{19}$B binding energy \cite{Gaud12}, may be used to determine the one- and
 two-neutron separation energies. These are plotted in Fig.~\ref{f:AME}, whereby the experimental results are
 compared with those derived from mass-surface extrapolations \cite{AME16,AME12}. The corresponding mass excesses
 are tabulated in Table~\ref{t:Mex}. As can be seen, the mass-surface extrapolations from the 2012 mass evaluation
 overbind $^{19,20,21}$B by $\sim1$--3~MeV. However, the more recent 2016 evaluation, which benefits from the
 $^{19}$B, $^{22}$C and $^{23}$N mass measurements \cite{Gaud12}, provides estimates for the mass excesses of
 $^{20,21}$B that are compatible with the present work. In this spirit, the present $^{20,21}$B masses will permit
 mass-surface extrapolations in this region to be made with improved precision and further from stability.

\header{Conclusions}
 In summary, using high-energy proton removal coupled with invariant-mass spectroscopy, the most neutron-rich boron
 isotopes to date have been observed for the first time. In the case of $^{20}$B a prominent resonance-like structure
 was observed at about 2.5~MeV above the one-neutron decay threshold that, guided by theoretical considerations, has
 been identified as the $1^-,2^-$ ground-state doublet, with energies $E_r=1.56\pm0.15$ and $2.50\pm0.09$~MeV.
 A weaker higher-lying peak was also observed at $4.86\pm0.25$~MeV ($E_x=3.30\pm0.29$~MeV). 
 The data acquired for $^{21}$B were consistent with the population of a resonance $2.47\pm0.19$~MeV above the
 two-neutron emission threshold, assigned to be the expected $3/2^-$ ground state.
 These results allowed the first determinations to be made of the ground-state masses of $^{20,21}$B, which are
 in agreement with the extrapolations of the most recent atomic-mass evaluations.
 In addition, $^{21}$B was found to exhibit direct two-neutron decay.  

 The identification and first spectroscopy of $^{20,21}$B presented here opens the way to the exploration of structure
 and correlations beyond the dripline below $^{24}$O. In particular, improvements in secondary-beam intensities and
 neutron detection should permit $n$-$n$ correlations in the decay of $^{21}$B to be investigated
 \cite{Dalitz,16Be,Comment-16Be} and its first excited state to be located. This, coupled with work underway to
 investigate the excited states of $^{22}$C, including the all important $2_1^+$ level \cite{Jones17,N=16,CC-C22},
 will provide direct insights into the $N=16$ shell closure beyond the neutron dripline as well as stringent tests
 of a new generation of {\em ab initio} and related theoretical models, including those incorporating explicitly
 the continuum.

\header{Acknowledgment}
 We wish to extend our thanks to the accelerator staff of the RIKEN Nishina Center for their efforts in
 delivering the intense $^{48}$Ca beam, and to C.~Yuan for the matrix elements of the YSOX interaction.
 N.L.A., F.D., J.G., F.M.M.\ and N.A.O.\ acknowledge partial support from the Franco-Japanese LIA-International
 Associated Laboratory for Nuclear Structure Problems as well as the French ANR-14-CE33-0022-02 EXPAND.
 A.N.\ and J.G.\ would like to acknowledge the JSPS Invitation fellowship program for long-term research in Japan
 at the Tokyo Institute of Technology and RIKEN, respectively.
 S.L.\ acknowledges the support provided by the short-term research International Associate Program of RIKEN,
 as well as the Tokyo Institute of Technology for the Foreign Graduate Student Invitation Program.
 This work was also supported in part by JSPS KAKENHI Grant No.~24740154 and 16H02179,
 MEXT KAKENHI Grant No.~24105005 and 18H05404, 
 the WCU (R32-2008-000-10155-0), the GPF (NRF-2011-0006492) programs of the NRF Korea, the HIC for FAIR,
 the CUSTIPEN (China-US Theory Institute for Physics with Exotic Nuclei)
 funded by the US Department of Energy, Office of Science under grant number DE-SC0009971,
 and the Office of Nuclear Physics under Awards No.~DE-SC0013365 (MSU) and No.~DE-SC0018083
 (NUCLEI SciDAC-4 Collaboration). 


\begin{thebibliography}{99}
 \bibitem{Nakamura} See, for example,
   T.~Nakamura, H.~Sakurai, H.~Watanabe, Prog.\ in Part.\ Nucl.\ Phys.\ {\bf97}, 53 (2018).
 \bibitem{Dripline} T.~Otsuka,           Phys.\ Scr.\ {\bf T152}, 014007 (2013).
 \bibitem{Otsuka10} T.~Otsuka \etal,     Phys.\ Rev.\ Lett.\ {\bf105}, 032501 (2010).
 \bibitem{IoI}      E.K.~Warburton, J.A.~Becker, B.A.~Brown, Phys.\ Rev.\ C {\bf41}, 1147 (1990).
 \bibitem{Stan04}   M.~Stanoiu \etal,    Phys.\ Rev.\ C {\bf69}, 034312 (2004).
 \bibitem{Hoffman}  C.R.~Hoffman \etal,  Phys.\ Lett.\ B {\bf672}, 672 (2009).
 \bibitem{BAB-WAR}  B.A.~Brown, W.A.~Richter,  Phys.\ Rev.\ C {\bf72}, 057301 (2005).
 \bibitem{abnit} See, for example, H. Hergert \etal, Phys.\ Rev.\ Lett.\ {\bf110}, 242501 (2013).
 \bibitem{GSM}      Y.~Jaganathen \etal,  Phys.\ Rev.\ C {\bf96}, 054316 (2017).
 \bibitem{Boron8}   T.~Minamisono \etal, Phys.\ Rev.\ Lett.\ {\bf69}, 2058 (1992).
 \bibitem{JL}       J.L.~Lecouey \etal,  Phys.\ Lett.\ B {\bf672}, 6 (2009).
 \bibitem{Spyrou10} A.~Spyrou \etal,     Phys.\ Lett.\ B {\bf683}, 129 (2010).
 \bibitem{Suzuki99} T.~Suzuki \etal,     Nucl.\ Phys.\ A {\bf658}, 313 (1999).
 \bibitem{Ozawa2003} A.Ozawa \etal,  Phys.\ Rev.\ C {\bf67}, 014610 (2003).
 \bibitem{Kondo16}  Y.~Kondo \etal,      Phys.\ Rev.\ Lett.\ {\bf116}, 102503 (2016).
 \bibitem{Togano16} Y.~Togano \etal,     Phys.\ Lett.\ B {\bf761}, 412 (2016).
 \bibitem{Kubo03}   T.~Kubo,               Nucl.\ Instrum.\ Methods Phys.\ Res., Sect.\ B {\bf204}, 97 (2003).
 \bibitem{Koba13}   T.~Kobayashi \etal,    Nucl.\ Instrum.\ Methods Phys.\ Res., Sect.\ B {\bf317}, 294 (2013).
 \bibitem{Kondo15}  T.~Nakamura, Y.~Kondo, Nucl.\ Instrum.\ Methods Phys.\ Res., Sect.\ B {\bf376}, 1 (2015).
 \bibitem{Sato-SAMURAI} H.~Sato \etal,     IEEE Trans.\ Applied Superconductivity {\bf23}, 4500308 (2013).
 \bibitem{Shimizu}  Y.~Shimizu \etal,      Nucl.\ Instrum.\ Methods Phys.\ Res., Sect.\ B {\bf317}, 739 (2013).
 \bibitem{Gaud12}   L.~Gaudefroy \etal,  Phys.\ Rev.\ Lett.\ {\bf109}, 202503 (2012).
 \bibitem{BW}       A.M.~Lane, R.G.~Thomas, Rev.\ Mod.\ Phys.\ {\bf30}, 257 (1958).
 \bibitem{MIX}      F.M.~Marqu\'es \etal, Phys.\ Lett.\ B {\bf476}, 219 (2000).
 \bibitem{Giacomo}  G.~Randisi \etal,     Phys.\ Rev.\ C {\bf89}, 034320 (2014).
 \bibitem{Supp-url} See Supplemental Material at http://link.aps.org/
                    supplemental/10.1103/PhysRevLett.121.262502
                    for the single-resonance fit of $^{20}$B and the three-body decay of $^{21}$B.
 \bibitem{crosstalk}T.~Nakamura, Y.~Kondo, Nucl.\ Instrum.\ Methods Phys.\ Res., Sect.\ B {\bf376}, 156 (2016).
 \bibitem{GSI-26O}  C.~Caesar \etal,      Phys.\ Rev.\ C {\bf88}, 034313 (2013).
 \bibitem{Dalitz}   F.M.~Marqu\'es \etal, Phys.\ Rev.\ C {\bf64}, 061301 (2001).
 \bibitem{SM}       E.~Caurier \etal, Rev.\ Mod.\ Phys.\ {\bf77}, 427 (2005), and references therein.
 \bibitem{Yuan12}   C.~Yuan \etal,        Phys.\ Rev.\ C {\bf85}, 064324 (2012).
 \bibitem{Sylvain}  S.~Leblond, PhD Thesis, Normandie Universit\'e (2015); S.~Leblond \etal, in preparation. 
 \bibitem{Nicolas}  N.~Michel \etal, in preparation.
 \bibitem{Tshoo12}  K.~Tshoo \etal,       Phys.\ Rev.\ Lett.\ {\bf109}, 022501 (2012).
 \bibitem{Jones17}  M.D.~Jones \etal,     Phys.\ Rev.\ C {\bf95}, 044323 (2017).
 \bibitem{AME16}    M.~Wang \etal,       Chinese Phys.\ C {\bf41}, 030003 (2017).
 \bibitem{AME12}    M.~Wang \etal,       Chinese Phys.\ C {\bf36}, 1603 (2012).
 \bibitem{16Be}     A.~Spyrou \etal,     Phys.\ Rev.\ Lett.\ {\bf108}, 102501 (2012).
 \bibitem{Comment-16Be} F.M.~Marqu\'es \etal,   Phys.\ Rev.\ Lett.\ {\bf109}, 239201 (2012).
 \bibitem{N=16}     M.~Stanoiu \etal,    Phys.\ Rev.\ C {\bf78}, 034315 (2004)  
 \bibitem{CC-C22}   G.R.~Jansen \etal,   Phys.\ Rev.\ Lett.\ {\bf113}, 142502 (2014).
\end{thebibliography}

\end{document}